\documentstyle[12pt,aasms4]{article}

\lefthead{Berger et al.}
\righthead{Discovery of 800 Hz QPO in 4U\,1608$-$52}

\begin{document}

\title{Discovery of 800 Hz QPO in 4U\,1608$-$52}

\author{M. Berger\altaffilmark{1}, M. van der Klis\altaffilmark{1},
J. van Paradijs\altaffilmark{1,}\altaffilmark{2},
W.H.G. Lewin\altaffilmark{3}, F. Lamb\altaffilmark{4},
B. Vaughan\altaffilmark{5}, E. Kuulkers\altaffilmark{6},
T. Augusteijn\altaffilmark{7}, W. Zhang\altaffilmark{8,9},
F.E. Marshall\altaffilmark{8}, J.H. Swank\altaffilmark{8},
I. Lapidus\altaffilmark{10}, J.C. Lochner\altaffilmark{8,9} and
T.E. Strohmayer\altaffilmark{8,9}}

\altaffiltext{1}{Astronomical Institute ``Anton Pannekoek'', University of Amsterdam and Center for High Energy Astrophysics, Kruislaan 403, 1098 SJ Amsterdam; michielb@astro.uva.nl, michiel@astro.uva.nl, jvp@astro.uva.nl}

\altaffiltext{2}{Physics Department, University of Alabama in Huntsville, Huntsville, AL 35899}

\altaffiltext{3}{Department of Physics and Center for Space Research, Massachusetts Institute of Technology, Cambridge, MA 02139; lewin@space.mit.edu}

\altaffiltext{4}{Physics Department, University of Illinois at
Urbana-Champaign, 1110 West Green Street, Urbana, IL 61801-3080;
f-lamb@uiuc.edu}

\altaffiltext{5}{Space Radiation Laboratory, California Institute of Technology, 220-47 Downs, Pasadena, CA 91125; brian@thor.srl.caltech.edu}

\altaffiltext{6}{ESA/ESTEC, Astrophysics Division, P.O.~Box 299, 2200 AG Noordwijk, The Netherlands; ekuulkers@astro.estec.esa.nl}

\altaffiltext{7}{European Southern Observatory, La Silla, Chile; tauguste@eso.org}

\altaffiltext{8}{NASA/GSFC, Laboratory for High Energy Astrophysics, Greenbelt, MD 20771; 
zhang@xancus10.gsfc.nasa.gov, marshall@rosserv.gsfc.nasa.gov, swank@lheavx.gsfc.nasa.gov, lochner@xeric.gsfc.nasa.gov, stroh@pcasrv1.gsfc.nasa.gov}

\altaffiltext{9}{also Universities Space Research Association}
\altaffiltext{10}{Institute of Astronomy, University of Cambridge, Maddingley Road, Cambridge CB 30HA, UK; lapidus@ast.cam.ac.uk}

\begin{abstract}

We present results of RXTE observations of the low-mass X-ray binary
and atoll source 4U\,1608$-$52 made over 9 days during the decline of an
X-ray intensity outburst in March 1996. A fast-timing analysis shows a
strong and narrow quasi periodic oscillation (QPO) peak at frequencies
between 850 and 890 Hz on March 3 and 6, and a broad peak around 690
Hz on March 9. Observations on March 12 show no significant signal.
On March 3, the X-ray spectrum of the QPO is quite hard; its strength
increases steadily from 5 \% at $\sim$2 to $\sim$20 \% at
$\sim$12~keV. The QPO frequency varies between 850 and 890 Hz on that
day, and the peak widens and its rms decreases with centroid frequency
in a way very similar to the well-known horizontal branch oscillations
(HBO) in Z-sources.  We apply the HBO beat frequency model to atoll
sources, and suggest that, whereas the model could produce QPOs at the
observed frequencies, the lack of correlation we observe between QPO
properties and X-ray count rate is hard to reconcile with this model.

\end{abstract}


\section{Introduction}

The low mass X-ray binary 4U\,1608$-$52 was first observed in 1971
(\cite{tan}).  It is the same source as the Norma burster, from which
the first X-ray bursts were discovered (\cite{belian}), independent
from the X-ray burst discovered by Grindlay et al. (1976) from
4U\,1820$-$30. 4U\,1608$-$52 is a soft X-ray transient, which shows
outbursts at intervals varying between $\sim 100$ days and several
years (see, e.g., \cite{lochner}).

Hasinger and Van der Klis (1989) classified 4U\,1608$-$52 as an atoll
source, based on the correlated X-ray spectral variability and
$\lesssim$ 10 Hz noise in the X-ray intensity that is characteristic
for this class of objects.  In the standard description of atoll
sources (e.g. \cite{vdk96}) the correlated changes in X-ray spectral
and timing properties are attributed to changes in mass accretion
rate.  At low accretion rates (island state of atoll sources) the
X-ray spectrum of 4U\,1608$-$52 contains a hard power law component
(up to about 100 keV) somewhat similar to the X-ray spectra of
black-hole candidates in their low state (\cite{mitsuda}). 
Yoshida et al. (1993)
found the power density spectrum of 4U\,1608$-$52 in its island state
has the shape characteristic of black hole candidates in the low
state, and reported 2--9 Hz quasi-periodic oscillations (QPO) from
this source, which is unusual in atoll sources.

In this paper we report the discovery of 650--850 Hz QPO from
4U\,1608$-$52. A preliminary announcement of this discovery was
already made by van Paradijs et al. (1996).  4U\,1608$-$52 is the
second atoll source showing 800 Hz QPO after 4U\,1728$-$34
(\cite{stroh1}). QPO at 800 and 1100 Hz have recently also been
reported for the Z-source Sco X-1 (van der Klis 1996a,b).

\section{Observations and analysis}

We observed 4U\,1608$-$52 using the proportional counter array (PCA)
on board of NASA's Rossi X-ray Timing Explorer (RXTE; \cite{xte})
during the decay of an X-ray outburst on March 3, 6, 9 and 12 of 1996.
Each observation had a duration of several satellite orbits, in each
of which the source was visible for about 3600 seconds, and occulted
by the Earth for about 1800 seconds.

The source count rate during the March 3 observation varied between
2910 and 3400 c/s (in the PCA energy range of 5--60 keV), on March 6
it had declined to 530--820 c/s, and on March 9 and 12 it was 610--730
and 460--710 c/s, respectively. No X-ray bursts were seen.  The
background count rate was about 90 c/s.  In all observations, data
were collected with a sampling rate of 8 kHz, except for the March 6
observation which used 16 kHz sampling. No spectral analysis of the
data was done yet, but the drop by a factor of 5 in X-ray intensity
could be due to a change from banana to island state (cf.~Fig.~1 of
\cite{mitsuda}).

We calculated power spectra of these data. We find that during most of
our observations these contain a clear QPO peak at high frequencies,
which we fitted with a constant level plus a Lorentzian. We corrected
the results of these fits for background and differential dead time
(\cite{vdk89}).  The reduced $\chi^2$ values of the fits were all
$\sim 1$.

\section{Results}

In the March 3 and 6 observations, a clear QPO peak can be seen in the
power spectrum (see Fig.~1 for an example).  In Figure 2, the
dynamical power spectra of these two observations are shown.  The
centroid frequency of the QPO varies between 850 and 890 Hz and the
peak is always narrow with Q values (centroid frequency divided by
full width at half maximum) of up to $\sim 200$.  During the first
orbit of the March 3 observation, the QPO frequency variations covered
the 850--890 Hz range in less than 1000 s.

We made fits to the QPO in contiguous segments of 100 seconds of the
March 3, orbit 1 data, where the QPO frequency changes rapidly.  The
rms amplitude of the QPO is between $\sim$6 and $\sim$8 \% and its
FWHM between 4 and 15 Hz.  In Figure 3, the rms amplitude and FWHM are
plotted as a function of QPO frequency.  As the QPO frequency
increases, the peak becomes weaker and broader.

In the second orbit of March 3 and in the March 6 data, the QPO
frequency changes much less but still significantly (Fig.~2). The QPO
strength, centroid and width values of orbit 2 of March 3 are
consistent with those of orbit 1. The March 6 peak widths are also
consistent with those of March 3 for the frequency range that they
have in common (850-870 Hz), but the fractional rms amplitude is
twice as high ($\sim$ 14 \%). In both observations, changes in QPO
frequency occur down to time scales of tens of seconds, as can be
seen in Fig.~2.

In the March 9 observation, no QPO are evident in the dynamical power
spectrum. Analyzing a power spectrum averaged over the entire
observation we find a broad peak with an amplitude of 13.9 $\pm$ 0.9
\% rms, a FWHM of 131 $\pm$ 19 Hz and a centroid frequency of 691
$\pm$ 6 Hz.  The FWHM of the QPO must be intrinsically
higher during this observation than during the earlier ones, or with this 
amplitude it would have been visible in the dynamical power spectrum.

On March 12, no $\gtrsim 100$ Hz QPO signal is detected in the
integrated power spectrum, but a very weak and broad band-limited
noise component can be seen at $\sim$90 Hz. It is uncertain whether or
not this feature is related to the $\sim$800 Hz QPO.

On March 6, there is a correlation between X-ray count rate and QPO
frequency. However, in the March 3 observation the changes
in QPO frequency are not well correlated to the count rate changes,
the drop in count rate by more than a factor 4 from March 3 to March 6
does not affect QPO frequency (or peak width), and in the March 9
observation, at nearly the same count rate as during March 6, the QPO
frequency is 150 Hz lower than before. It is therefore possible that
the correlation on March 6 is coincidence.

We looked for the presence of a harmonic to the QPO peak at twice
the centroid frequency in the March 3, orbit 2 data; the 3$\sigma$
upper limit to such a component is 1.7 \% rms, corresponding to an
amplitude ratio of $<4.1$

We measured the photon energy dependence of the QPO in the second
orbit of the March 3 data by dividing the data into 8 contiguous
energy bands centered at 5.0, 7.1, 9.4, 11.8, 14.8, 18.2, 22.2 and
27.2 keV. We used this data because of the strength and stable
frequency of the QPO, making it easy to construct an average power
spectrum from each band. We analyzed the power spectra as described
above.  The rms amplitude of the QPO is plotted as a function of
energy in Figure 4.  The QPO relative amplitude increases up to about
12 keV.

\section{Discussion}

In several respects, the $\sim$800 Hz QPO we found in 4U\,1608$-$52 are very
similar to those in 4U\,1728$-$34 (\cite{stroh1}). The frequency
ranges, coherencies, peak shapes and strengths are all approximately
the same. An important difference of the $\sim$800 Hz QPO in these
two atoll sources with those at 800 and 1100 Hz in the Z~source
Sco~X-1 (van der Klis et al., 1996a,b) is the much lower ($\sim$
factor 10) rms amplitude of the QPO in the latter. This could be due to
scattering in circumstellar material or additional, unmodulated flux
in the case of the near-Eddington accretion thought to characterize
Z~sources.

There are two lines of reasoning that suggest that the 800 Hz QPO
in 4U\,1608$-$52 may be due to a beat between Keplerian disk rotation
and neutron star spin. However, our data also provide a strong argument
against this interpretation.

First, the correlation between the various QPO properties during March
3, orbit 1 is very reminiscent of the horizontal branch oscillations
(HBO) as seen in the Z source GX\,5$-$1 (\cite{vdk85}, cf.~their
Fig.~3).  When the QPO frequency increases, the peak width increases
and the rms amplitude goes down.  The rms vs.~photon energy spectrum
of the QPO is very similar to that of GX\,5$-$1 (\cite{lewin}), which
increases from ~4 to ~15 \% rms between 2 and 15 keV.  The width of
the $\sim$40 Hz QPO of GX5-1 is between 5 and 12 Hz (\cite{erik}),
very similar to the 5-15 Hz width of the $\sim$800 Hz QPO of
4U\,1608$-$52. A small difference is that in GX5-1 the rms amplitude
of the QPO is between 2 and 6 \%, which is slightly lower than the
5-15 \% seen in 4U\,1608$-$52.

We note that many of these similarities could also be due to the 800
Hz QPO in 4U\,1608$-$52 being directly due to the inner disk frequency
(not the beat frequency).

The beat frequency model (\cite{alpar}, \cite{lamb}) for HBO in
Z~sources could, with an assumed neutron star spin rate of a few 100
Hz and with the lower magnetic field strengths expected for atoll
sources as compared to Z sources (\cite{hk89}) very easily produce an
800 Hz beat frequency.

Second, Strohmayer et al. (1996b) have proposed that the 800 Hz QPO in
4U\,1728$-$34 can be interpreted in terms of a beat frequency between
the Keplerian disk frequency and a neutron star spin rate of 363 Hz.
The report that five X-ray bursts of that source show a relatively
coherent signal at 363 Hz, and the difference in frequency between the
800 Hz QPO peak with another peak present in their data remains
consistent with being equal to 363 Hz across the observations.

However, in spite of these arguments, the absence of a correlation
between X-ray intensity on the one hand and QPO frequency and rms
amplitude on the other hand poses a major problem for a beat frequency
model interpretation.  The QPO frequency remains constant near 850 Hz
between 3 and 6 March while the count rate drops by more than a
factor 4, from $\sim$3200 to $\sim$600 counts/sec (Sect.~2).

The beat frequency model predicts that if the QPO frequency remains
constant, the mass flow through the inner edge of the disk should
remain constant as well. Our data therefore are inconsistent with a
beat frequency model interpretation if all accretion takes place via
the inner edge of the disk. We note that the presence of a
hypothetical additional (non-disk) mass flow component contributing
more than 75\% of the total flux on March 3 and much weaker on March 6
can not easily resolve this discrepancy.  The rms amplitude of the
QPO only increases by a factor 2 from March 3 to 6 while the X-ray
intensity drops by a factor of more than 4, whereas a similar
fractional change would be predicted in this explanation. 
Only by invoking rather large and entirely ad-hoc changes in the beaming or bolometric correction could the model be maintained.
The 150 Hz drop
in QPO from March 6 to March 9 without a change in count rate presents
similar difficulties.

Further observations of 4U\,1608$-$52 during other outbursts, and
study of its X-ray bursts are required to shed further light on the
relation of the 800 Hz QPO in 4U\,1608$-$52 with those in
4U\,1728$-$34, and with the QPO in Sco X-1.

\placefigure{fig1}
\placefigure{fig2}
\placefigure{fig3}
\placefigure{fig4}

\acknowledgments

This work was supported in part by the Netherlands Organization for
Scientific Research (NWO) under grant PGS 78-277 and by the
Netherlands Foundation for Research in Astronomy (ASTRON) under grant
781-76-017. WHGL and JVP acknowledge support from the National
Aeronautics and Space Administration.

\clearpage

\figcaption{Leahy-normalized power spectrum of 4U\,1608$-$52, from data of March 6. No corrections for counting statistics and dead time have been made for this figure.\label{fig1}}

\figcaption{Dynamical power spectra of March 3, orbit 1 and 2 (top
panel) and March 6, orbit 1 (bottom panel). \label{fig2}}

\figcaption{Properties of the QPO of 4U\,1608$-$52.  In the left panel the
relation between QPO centroid frequency and FWHM is given, the right
panel shows the relation between QPO frequency and fractional rms
amplitude. The errors correspond to $\Delta\chi^2=1$ in
the fit.\label{fig3}}

\figcaption{The rms vs. photon energy spectrum of the QPO. Plotted is
the fractional rms amplitude of the fitted Lorentzian to the QPO. \label{fig4}}

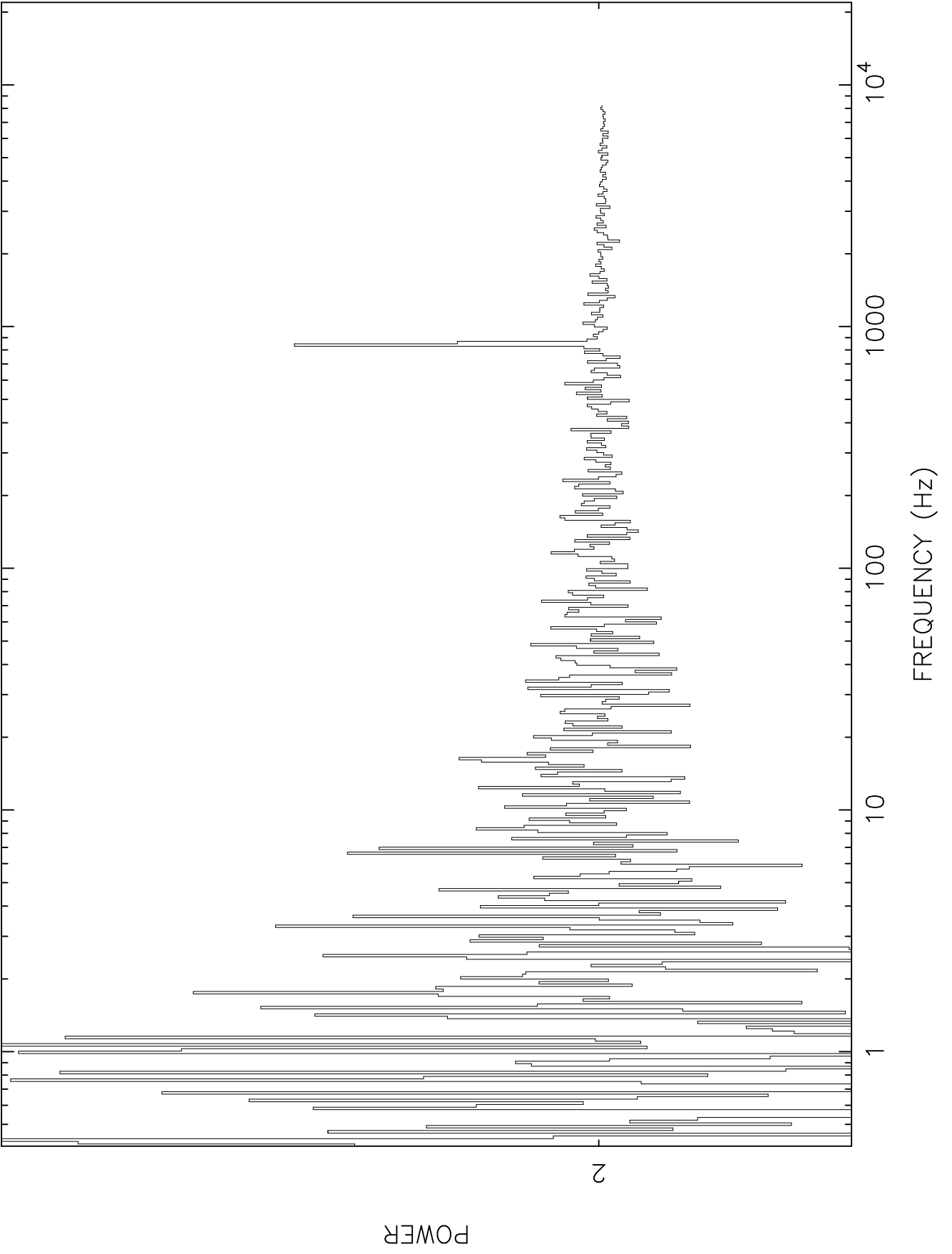
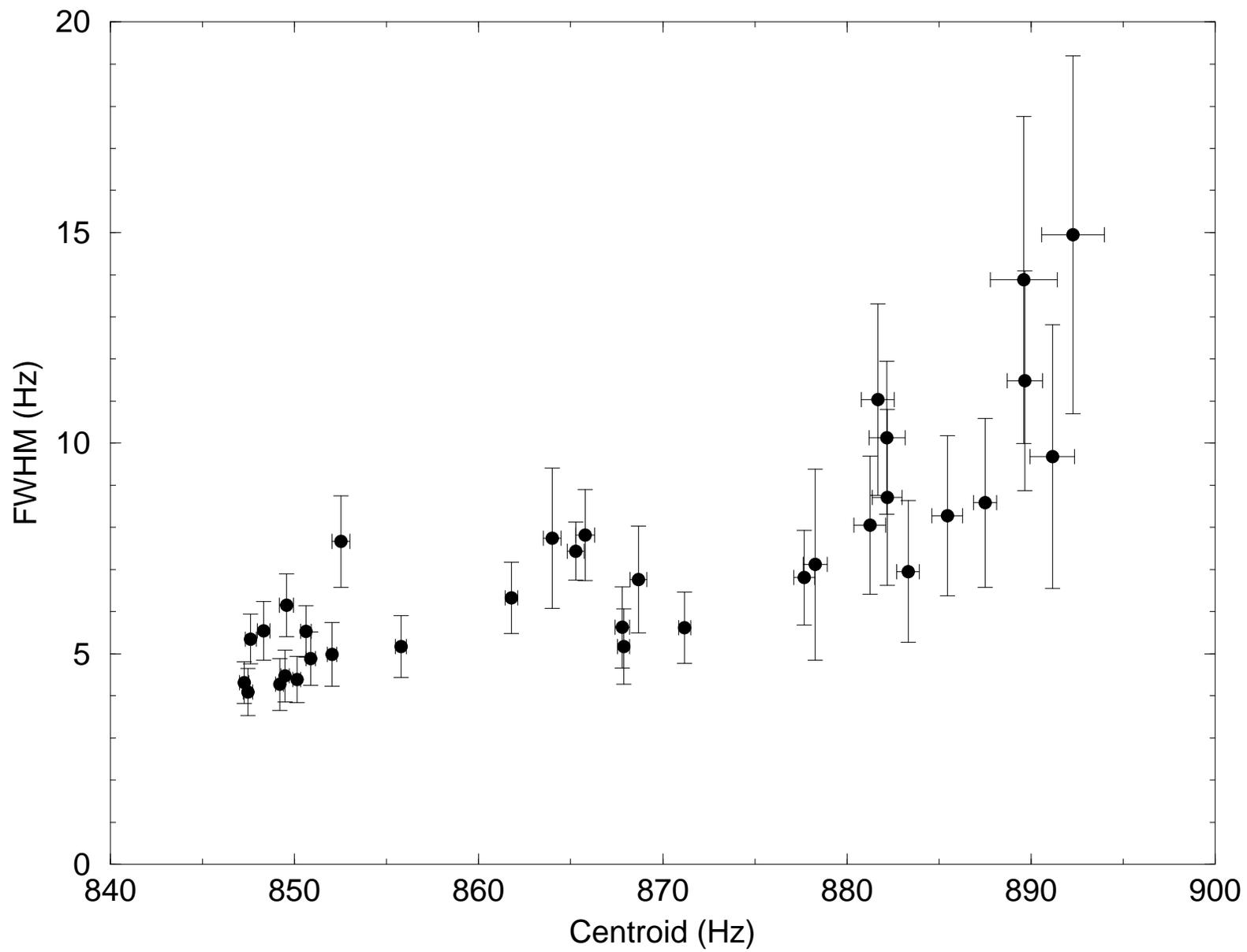
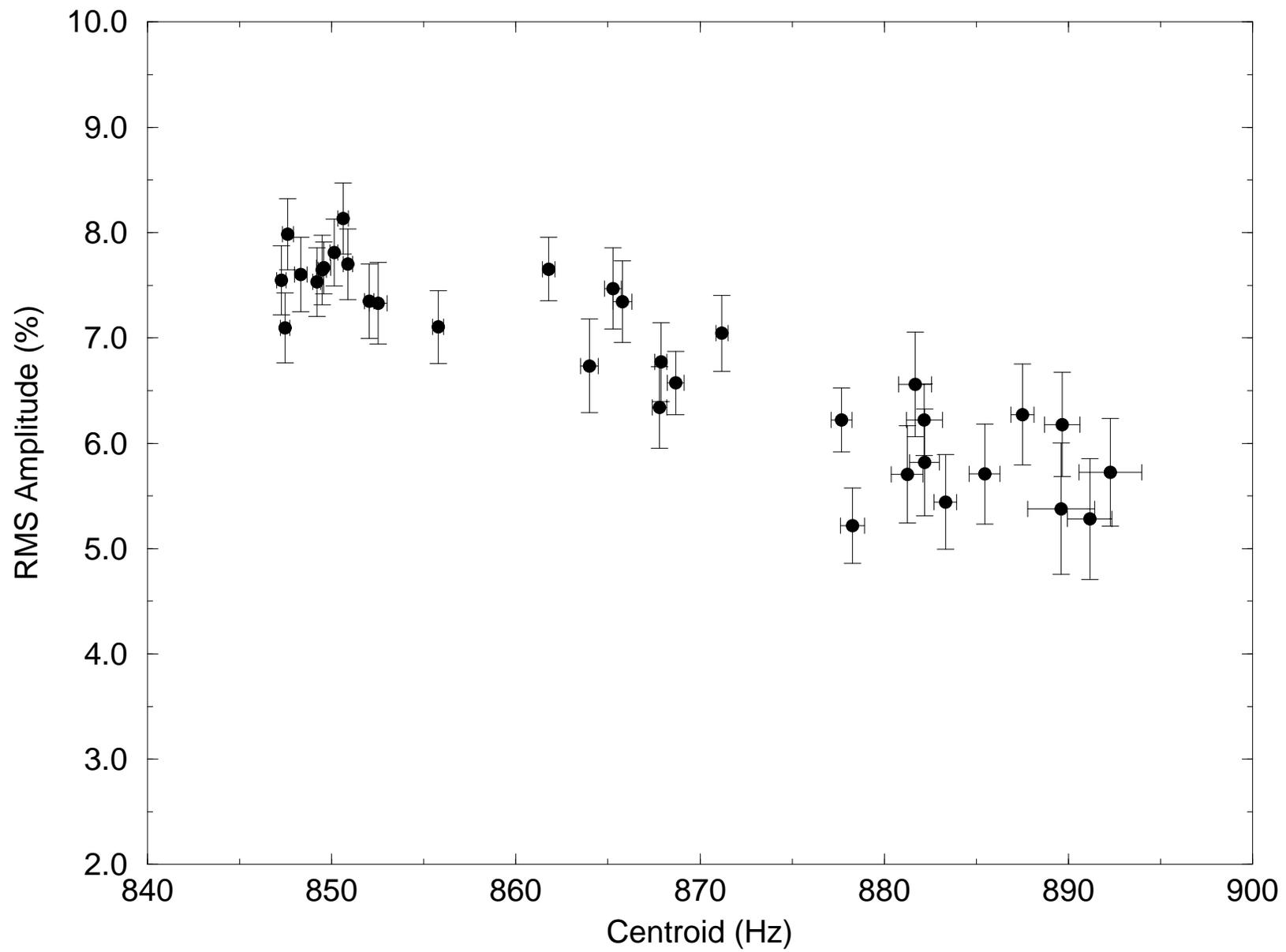
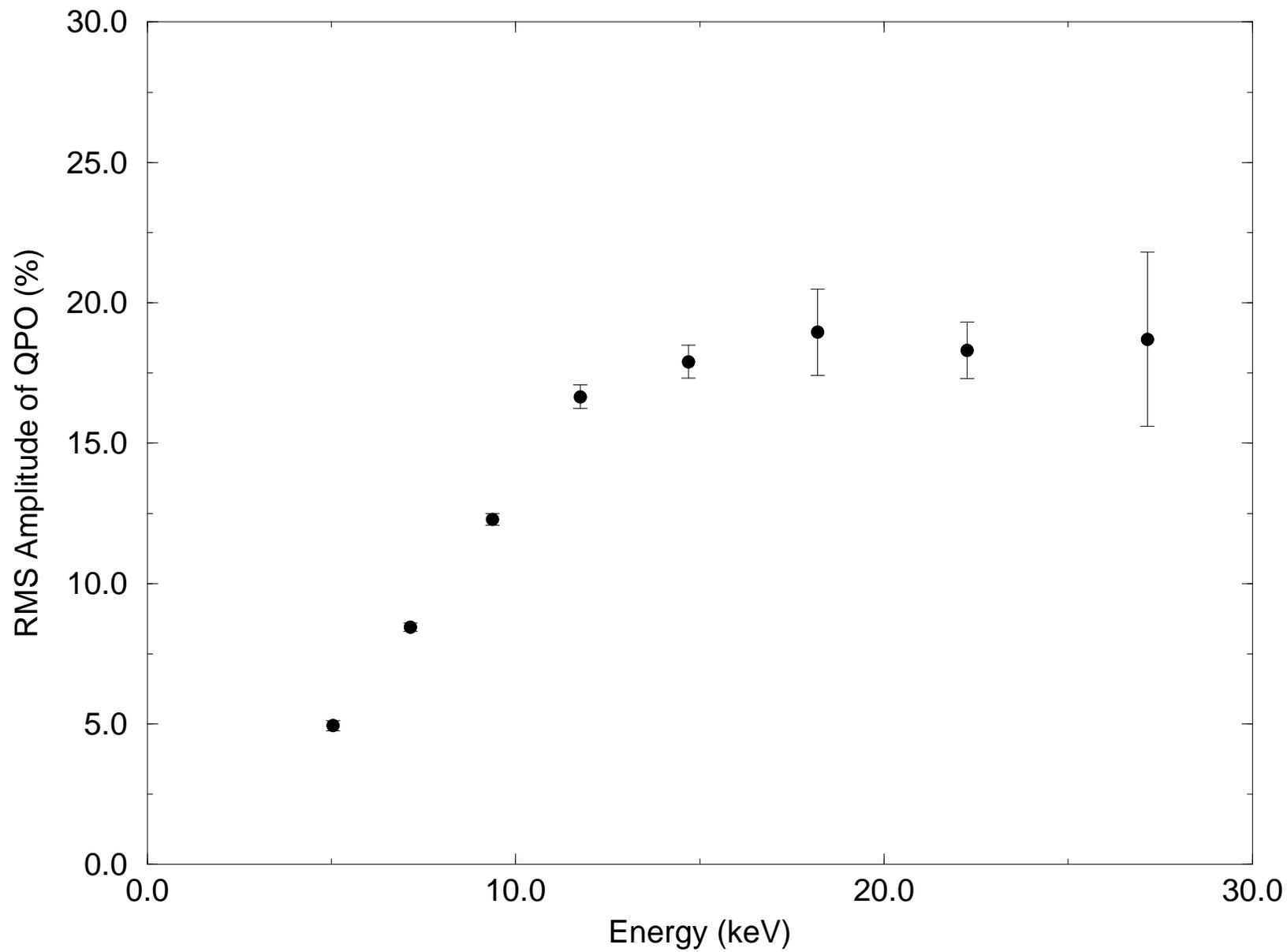
\end{document}